\documentclass[prb,twocolumn,superscriptaddress,floatfix]{revtex4}
\usepackage{amsmath,amssymb,epsfig,psfig}
\begin{document}

\title{Orbital state and magnetic properties of LiV$_2$O$_4$}

\author{I.A.~Nekrasov}
\affiliation{Institute of Metal Physics, Russian Academy of Sciences-Ural
Division, 620219 Yekaterinburg GSP-170, Russia}
\affiliation{Theoretical Physics III, Center for Electronic Correlations and
Magnetism, Institute for Physics,
University of Augsburg, 86135 Augsburg, Germany}
\author{Z.V.~Pchelkina}
\affiliation{Institute of Metal Physics, Russian Academy of Sciences-Ural
Division, 620219 Yekaterinburg GSP-170, Russia}
\affiliation{Theoretical Physics III, Center for Electronic Correlations and
Magnetism, Institute for Physics,
 University of Augsburg, 86135 Augsburg, Germany}
\affiliation{Department of Theoretical Physics and Applied Mathematics,
Ural State Technical University, 620002 Yekaterinburg Mira 19, Russia}
\author{G.~Keller}
\affiliation{Theoretical Physics III, Center for Electronic Correlations and
Magnetism, Institute for Physics,
 University of Augsburg, 86135 Augsburg, Germany}
\author{Th.~Pruschke}
\affiliation{Theoretical Physics III, Center for Electronic Correlations and
Magnetism, Institute for Physics,
 University of Augsburg, 86135 Augsburg, Germany}
\author{K.~Held}
\affiliation{Max Planck Institute for Solid State Research, Heisenbergstr.~1,
70569 Stuttgart, Germany}
\author{A.~Krimmel}
\affiliation{Experimental Physics V,
Center for Electronic Correlations and Magnetism, Institute for Physics,
University of Augsburg, 86135 Augsburg, Germany}
\author{D.~Vollhardt}
\affiliation{Theoretical Physics III, Center for Electronic Correlations and
Magnetism, Institute for Physics,
 University of Augsburg, 86135 Augsburg, Germany}
\author{V.I.~Anisimov}
\affiliation{Institute of Metal Physics, Russian Academy of Sciences-Ural
Division, 620219 Yekaterinburg GSP-170, Russia}
\affiliation{Theoretical Physics III, Center for Electronic Correlations and
Magnetism, Institute for Physics,
 University of Augsburg, 86135 Augsburg, Germany}

\date{\today}

\begin{abstract}

LiV$_2$O$_4$ is one of the most puzzling compounds among transition metal oxides
because of its heavy fermion like behavior at low temperatures.
In this paper we present results for the orbital state and 
magnetic properties of LiV$_2$O$_4$ obtained from a combination of 
density functional theory within the local density approximation
and dynamical mean-field theory (DMFT). The DMFT equations are
solved by quantum Monte Carlo simulations.
The trigonal crystal field splits the V $3d$ orbitals
such that the 
 a$_{1g}$   and  e$_{g}^\pi$ orbitals cross the Fermi level,
with the former being slightly lower
in energy and narrower in bandwidth.
In this situation, the $d$-$d$ Coulomb interaction leads to an almost localization
of one electron per V ion in the a$_{1g}$ orbital, while the e$_{g}^\pi$
orbitals form relatively broad bands with 1/8 filling.
The theoretical high-temperature paramagnetic susceptibility $\chi(T)$ follows
a Curie-Weiss law with an effective paramagnetic moment $p_{eff}$=1.65
in agreement with the experimental results.
\end{abstract}

\maketitle

\section{Introduction}

Heavy-fermion (HF) materials are typically intermetallic compounds containing
Ce, U, or Yb atoms. They are characterized by extraordinarily strongly
renormalized effective masses, $m^*\approx100-1000
~m_e$,\cite{Anders75,Stewart84,Aeppli92} as 
inferred from the electronic specific heat coefficient $\gamma(T)\equiv
C_e(T)/T$ at low temperature. They also show an apparent local moment 
paramagnetic behavior with a strongly enhanced spin susceptibility $\chi$
at low temperatures. 
The discovery by Kondo~{\it et~al.}~\cite{Kondo7}
of the HF behavior in LiV$_2$O$_4$ with a Kondo or spin fluctuation temperature 
T$_K\sim\,$28 K has significant importance, because this is the first 
{\it d}-electron system that shows HF characteristics. Kondo~{\it et~al.}
reported a large electronic specific heat coefficient
$\gamma\approx 0.42\,$J/(mol K$^2$)
at~1~K, which is much larger than those of other metallic transition metal compounds
like,
e.g., Y$_{1-x}$Sc$_x$Mn$_2$ [$\gamma\lesssim$ 0.2 J/(mol K$^2$)]\cite{Ballou96} 
and V$_{2-y}$O$_3$ [$\gamma\lesssim$0.07~J/(mol K$^2$)].\cite{Carter93}
Also a crossover from the local moment to a renormalized
Fermi-liquid behavior was observed with decreasing temperature. 
Recently Urano~{\it et~al.}~\cite{Urano2000}
reported that the electrical resistivity $\rho$ of single crystals exhibits a T$^2$
temperature dependence $\rho=\rho_0+AT^2$ with an enormous A, which like in conventional HF 
systems scales with $\gamma^2$.\cite{kawo}
In the temperature range $50$--$1000\,$~K the experimental magnetic susceptibility
follows a Curie-Weiss law with a negativ Curie-Weiss temperature,
which indicates a weak antiferromagnetic (AF) V-V spin interaction
(see later in the text).
No magnetic ordering was observed down~to~0.02~K.\cite{Johnston99}

These unexpected phenomena entailed numerous experiments, which
confirmed the HF behavior 
of LiV$_2$O$_4$ in a variety of physical quantities: 
Johnston~{\it et~al.}~\cite{Johnston99} carried out specific
heat and thermal expansion measurements. Kondo~{\it et~al.}~\cite{Kondo99} described
the synthesis, characterization and magnetic susceptibility versus temperature.
Onoda~{\it et~al.}~\cite{Onoda97} explored spin fluctuations and transport 
in the spinel systems Li$_x$Mg$_{1-x}$V$_2$O$_4$ and Li$_x$Zn$_{1-x}$V$_2$O$_4$ 
through measurements of x-ray diffraction, electrical resistivity, thermoelectric power, 
magnetization, and nuclear magnetic resonance (NMR). 
The electron-spin resonance (ESR) and the magnetic 
susceptibility in pure and doped LiV$_2$O$_4$ were measured
by Lohmann~{\it et~al.}.\cite{Lohmann}
Photoemission studies of the hole doped Mott insulator
Li$_{1-x}$Zn$_x$V$_2$O$_4$ were done by Fujimori~{\it et~al.}.\cite{Fujimori88}
A series of $^7$Li-NMR experiments were carried out by
Fujiwara~{\it et~al.}~\cite{Fujiwara98} for LiV$_2$O$_4$
and also for Li$_{1-x}$Zn$_x$V$_2$O$_4$.\cite{Fujiwara99}
The Knight shift, spin susceptibility and
relaxation times were determined from $^7$Li-NMR experiments  by Mahajan~{\it et~al.}.\cite{Mahajan98}
Krimmel~{\it et~al.}~\cite{Krimmel99} presented results for the magnetic relaxation 
of LiV$_2$O$_4$ obtained by means of quasi-elastic neutron scattering. 
Trinkl~{\it et~al.}~\cite{Trinkl00} investigated spin-glass behavior in 
Li$_{1-x}$Zn$_x$V$_2$O$_4$. Urano~{\it et~al.}~\cite{Urano2000}
experimentally observed results for C(T),  $\chi$(T), resistivity $\rho$(T) and
Hall coefficient R$_H$(T) for single crystal samples. Recently,
Lee~{\it et~al.}~\cite{Lee01} performed inelastic neutron scattering measurements and
Fujiwara~{\it et~al.}~\cite{Fujiwara02} studied the spin dynamics 
under high pressure. A review of various experiments 
and theoretical research is collected in the work of Johnston.\cite{Johnston}

Let us summarize the experimental efforts concerning the measurements of
the spin susceptibility $\chi$(T). As mentioned before,
a temperature dependent spin susceptibility $\chi$(T) is observed
in the temperature range from 50 to 1000~K, which fits
well to the Curie-Weiss law\cite{Kondo7,Urano2000,Johnston99,%
Hayakawa,Kondo99,Onoda97,Lohmann,%
Fujiwara98,Fujiwara99,Mahajan98,Trinkl00,Lee01,Johnston,Takagi87,%
Nakajima91,Johnston95,Ueda97,Kessler71,Chamberland86,Brando02,Miyoshi02}
$\chi$(T)=$\chi_0$+C/(T-$\theta$).
However, depending on the quality of samples and experimental techniques,
the values of the Curie constant  C obtained from fitting the
experimental data are in the
range from 0.329--0.468~cm$^3$K/mol.\cite{Mahajan98,Kessler71}
More important and of direct physical importance is the value of
the effective paramagnetic moment $p_{eff}$, which is defined via the
Curie constant as $p^2_{eff}:=3Ck_B/(N_A\mu^2_B)$.
The extracted value $p_{eff}$ lies in the range between 1.62 and
2.1.\cite{Lee01,Miyoshi02}
The Curie-Weiss temperature obtained in those experiments is also
slightly different and encompasses the interval $\theta=-20\ldots-60$~K,
which indicates a weak antiferromagnetic (AF) V-V spin interaction.
If one further assumes that the effective spin value is S=1/2,
an experimental estimation of the Land\'e factor $g^2=p^2_{eff}/[S(S+1)]$ leads to
values between 1.87and 2.23.\cite{Mahajan98,Kessler71}

From a theoretical point of view, LiV$_2$O$_4$ has been intensively studied by
standard band structure calculations by various groups using different
implementations of density functional theory~\cite{dft}
within the local density approximation (DFT/LDA).\cite{lda}
Anisimov~{\it et~al.}~\cite{Anisimov99} investigated the possibility of
localization of the $d$-states in LiV$_2$O$_4$ within the linearized muffin-tin
orbitals (LMTO) basis~\cite{LMTO} supplemented by the LDA+U method.\cite{ldau}
The electronic structure was furthermore studied by 
Eyert~{\it et~al.}~\cite{Eyert99} with the scalar-relativistic augmented
spherical wave (ASW) basis,\cite{asw} while Matsuno~{\it et~al.}~\cite{Mattheiss}
used a full-potential, scalar-relativistic implementation~\cite{Mattheiss86}
of the linear augmented plane wave (LAPW)
approach~\cite{Andersen75} for the band structure calculation and furthermore applied
a simple tight-binding (TB) model.\cite{Mattheiss82}
Singh~{\it et~al.}~\cite{Singh99} also 
used the full potential LAPW~\cite{Singh94,Blaha,Wei}
for calculating the band structure and a TB-LMTO method~\cite{LMTO} to analyze
the band symmetry. 

A relative comparison of LiV$_2$O$_4$ and lithium titanate (LiTi$_2$O$_4$) with original
HF systems was done by Varma~\cite{Varma99} to provide some qualitative understanding
of these compounds.
The two-band Hubbard model in the slave-boson mean-field approximation was applied
to LiV$_2$O$_4$ by Kusunose~{\it et~al.}~\cite{Kusunose00} to investigate the
evolution of bands due to the Coulomb interactions.
Hopkinson~{\it et~al.} presented a simple two-band model~\cite{Hopkinson} 
and a $t$-$J$ model with a strong Hund's coupling for the 
$d$-electrons~\cite{Hopkinson312} to find evidence for
a two-stage screening in LiV$_2$O$_4$.
Fujimoto investigated the Hubbard chains network model on corner-sharing tetrahedra
as a possible microscopic model for the HF behavior in LiV$_2$O$_4$.\cite{Fujimoto}
The competition between the Kondo effect and frustrating exchange 
interactions in a Kondo-lattice model within a large-{\bf{\it N}} approach for the
spin liquid together with dynamical mean-field theory
were studied by Burdin~{\it et~al.}.\cite{Burdin}

Despite all this theoretical effort,
there does not yet exist an undisputed microscopic explanation of
the HF behavior of LiV$_2$O$_4$ at low temperatures.
It has been attempted to explain the low-temperature properties
of LiV$_2$O$_4$  by a mechanism analogous to one for systems with local moments
on a pyrochlor lattice,
which are frustrated with respect to the antiferromagnetic interactions.
A novel attempt to provide a
microscopic model based on material properties of LiV$_2$O$_4$, which does not
evoke the idea of frustration was suggested
by Anisimov~{\it et~al.}.\cite{Anisimov99}
Their basic idea is a separation of the electrons on partially filled t$_{2g}$ orbitals
into localized ones, forming local moments, and delocalized ones, producing
a partially filled metallic band. The hybridization between those two subsets
of electrons, like in $f$-electrons materials, can give rise to heavy-fermion effects.
However, in this work the conclusions were based on an LDA+U
calculation,\cite{ldau}
which is essentially a static mean-field approximation and surely too crude to
give a proper description of LiV$_2$O$_4$.

The present work reports on the investigation of the orbital state and the magnetic
properties of LiV$_2$O$_4$ using the novel LDA+DMFT(QMC) approach. This material
specific theory extracts informations about the non-interacting band structure
from a DFT/LDA calculation, while treating the local Coulomb interactions
via the dynamical mean-field theory (DMFT), which is a well-established
non-perturbative approach to study localization effects in strongly correlated
materials. The resulting DMFT equations are then solved by quantum
Monte-Carlo simulations.
The main goal of this study is to further explore the idea of a
separation of the $d$-electrons in LiV$_2$O$_4$ into two subsets,
localized and itinerant as proposed in Ref.~\onlinecite{Anisimov99}. 
The rest of the paper is divided into two parts:
In the first we discuss standard DFT/LDA
results (section~\ref{dftlda_res}) and then present our LDA+DMFT(QMC) results
in the second (section~\ref{ldadmft_res}).
A short summary concludes the paper.

\section{LiV$_2$O$_4$: DFT/LDA results}
\label{dftlda_res}

\subsection{Crystal structure and $d$ orbital splitting}
\label{crystal}

LiV$_2$O$_4$ has the fcc normal-spinel
structure with non-symmorphic space group $Fd3m$ and was first  
synthesized by Reuter and Jaskowsky in 1960.\cite{Reuter}
\begin{figure}[htb]
\centering
\epsfig{file=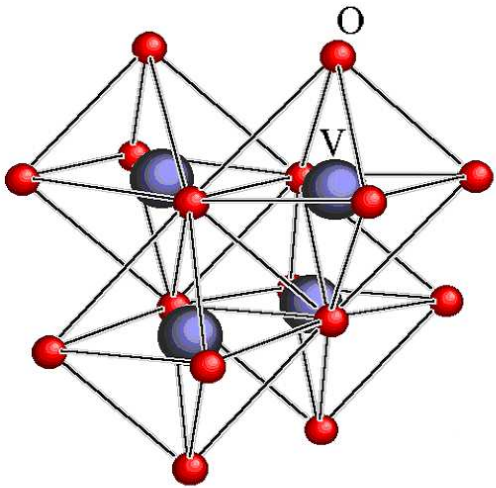,width=0.47\textwidth}
\caption{The normal-spinel crystal structure of LiV$_2$O$_4$ is formed 
by the oxygen edge-shared octahedra with V atoms at the centers. V: large
dark ions; O: small dark ions.} 
\label{octahedra}
\end{figure}
The Li ions are tetrahedrally
coordinated by oxygens, while the V sites are surrounded by a slightly distorted
edge-shared octahedral array of oxygens~(Fig. \ref{octahedra}).
The corresponding unit cell of the face-centered cubic lattice contains 
two LiV$_2$O$_4$ formula units (14~atoms) with four V atoms, which form a 
\begin{figure}[htb]
\centering
\epsfig{file=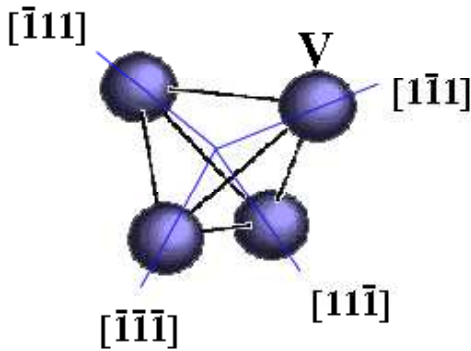,width=0.47\textwidth}
\caption{Tetrahedron formed by four V atoms (large dark spheres)
in the spinel unit cell and the trigonal axes for each V atom.} 
\label{tetra}
\end{figure}
tetrahedron~(Fig.~\ref{tetra}).
The LiV$_2$ substructure is the same as the C15 structure AB$_2$,
where the local moments at the B sites are highly frustrated.\cite{Wada89}
The observed
lattice constant of LiV$_2$O$_4$ is 8.22672 \AA~at 4 K.\cite{Kondo79}
The eight oxygen atoms in the primitive cell are situated at the
32{\it e}-type sites, at
positions which are determined by the internal-position parameter
x=0.2611 in units of the lattice 
constant.

The total space group of the crystal is cubic but the local point group symmetry
of the V ion crystallographic position is trigonal $D_{3d}$. The different trigonal axes
of every V atom in the unit cell are directed towards the center of the
tetrahedron~(Fig. \ref{tetra}).
Since the formal oxidation state of V is non-integer V$^{3.5+}$ (configuration $d^{1.5}$) 
and the V ions are crystallographically equivalent, LiV$_2$O$_4$ must be metallic as it is 
observed.\cite{Reuter}

The octahedral crystal field at the V sites in the spinel structure 
\begin{figure}
\centering
\epsfig{file=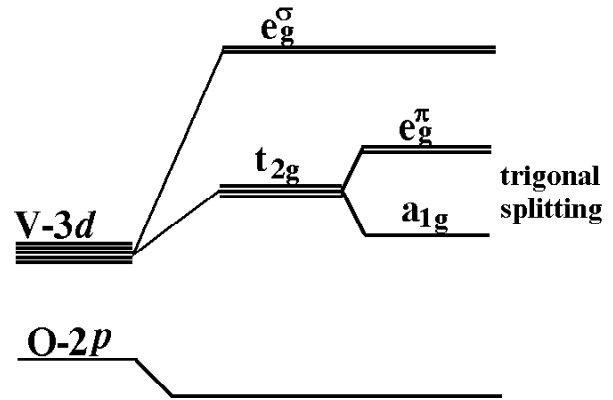,width=0.47\textwidth}
\caption{Schematic picture of the electronic level  distribution
caused by the strong $p$-$d$ hybridization
in the trigonally distorted octahedral crystal field.
The notations t$_{2g}$ and e$_g^\sigma$ correspond to the irreducible
representations of the cubic $O_h$ group, a$_{1g}$ and e$_g^\pi$
to the irreducible
representations of the $D_{3d}$ group.} 
\label{splitting}
\end{figure}
splits the V 3{\it d} bands into three degenerate and
partially filled t$_{2g}$ bands and two empty e$_g^\sigma$ bands.
The Fermi level lies within the t$_{2g}$ complex, thus 
the transport properties of LiV$_2$O$_4$ are solely associated with the t$_{2g}$ bands. 
The trigonal symmetry of the V ions splits the cubic t$_{2g}$ levels
into one  a$_{1g}$ and two degenerate
e$_{g}^\pi$ levels (Fig.~\ref{splitting}). However, this splitting is not large
enough to separate the t$_{2g}$ band into two subbands.\cite{Mattheiss}

\subsection{DFT/LDA band structure}
\label{dftlda_bands}

Based on DFT/LDA~\cite{dft,lda}
within the LMTO method,\cite{LMTO}
we performed first-principle calculations of the electronic structure
of LiV$_2$O$_4$. 
\begin{figure}[htb]
\centering
\epsfig{file=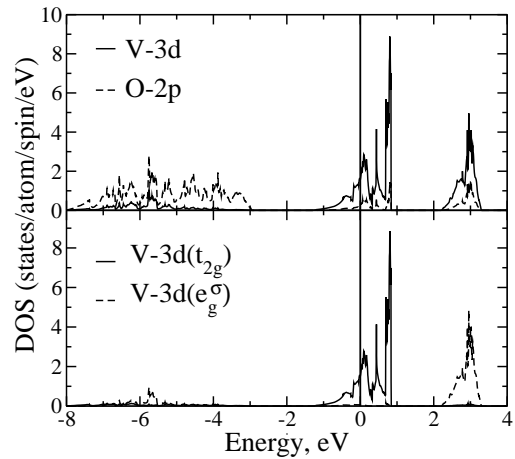,angle=270,width=0.47\textwidth}
\caption{DOS of LiV$_2$O$_4$ calculated with the LDA-LMTO method. Upper figure:~
V-3$d$ (full line) and O-2$p$ (dashed line) DOS; lower figure: partial V-3$d$(t$_{2g}$) 
(full line) and V-3$d$(e$_{g}^\sigma$) (dashed line) DOS.
The Fermi level corresponds to zero energy.} 
\label{dos_V_O}
\end{figure}
The radii of the muffin-tin spheres were R$_{\rm Li}$=2.00 a.u., R$_{\rm V}$=2.05 a.u.,
and R$_{\rm O}$=1.67 a.u. The resulting densities of states (DOS) are shown in 
Figs.~\ref{dos_V_O}~and~\ref{dos_V_t2g}. In~Fig.~\ref{dos_V_O},
there are three well-separated sets of bands:
\begin{figure}[htb]
\centering
\epsfig{file=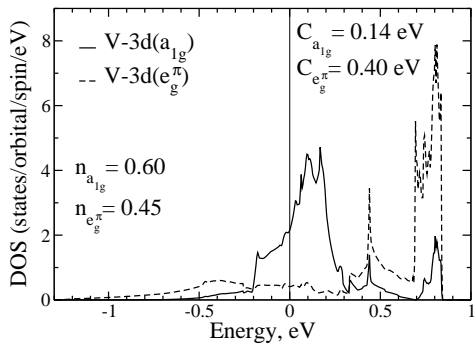,angle=270,width=0.4\textwidth}
\caption{Partial V-3$d$(t$_{2g}$) DOS of LiV$_2$O$_4$ calculated with
the LDA-LMTO method:
a$_{1g}$ (full line) and e$_g^\pi$ (dashed line) projected DOS.
The center of gravity of the a$_{1g}$ orbital is $c_{a_{1g}}\!=\,$0.14~eV;
that of the e$_{g}^\pi$ orbital is $c_{e_{g}^{\pi}}\!=\,$0.40~eV.
The LDA-calculated occupancies of the a$_{1g}$ and e$_{g}^\pi$ orbitals,
n$_{a_{1g}}$=0.60 and n$_{e_{g}^\pi}$=0.45, are nearly the same.
The Fermi level corresponds to zero energy.} 
\label{dos_V_t2g}
\end{figure}
completely filled O-2$p$-bands, partially filled t$_{2g}$ bands, and empty
e$_g^\sigma$ bands.
The bands in the energy range from -8 eV to -3 eV originate mainly from O-2$p$
states and have only 
small admixtures from V-3$d$ states. The upper two groups of bands,
which extend from -1.0 eV 
to 0.8 eV and from 2.3 eV to 3.2 eV, are predominantly derived from the V-3d states.
Although additional small O-2$p$ contributions are apparent in this energy range,
the $p$-$d$
hybridization is much reduced compared to other early transition metal
oxides.\cite{Eyert98}
Due to the  crystal field of the slightly distorted octahedral coordination
of the V atoms by the oxygen atoms,  a clear separation of the 3$d$
t$_{2g}$ and e$_g^\sigma$ groups of bands is visible in ~Fig.~\ref{dos_V_O}. 
Whereas the former states appear exclusively around the Fermi energy, the e$_g^\sigma$
states prevail at higher energies.
Contributions of the V 3{\it d} states to the oxygen-derived
bands originate almost exclusively from the e$_g^\sigma$ states,
which form $\sigma$-bonds and experience a strong overlap
with the O-2{\it p} states. In contrast, the  t$_{2g}$ orbitals, which give rise
to $\pi$ bonds, yield only a negligible contribution in this energy range.
In addition to $p$-$d$ bonds, the t$_{2g}$ states experience
strong $\sigma$-type overlap
with the t$_{2g}$ orbitals at neighboring vanadium sites of the fcc sublattice. 
Hence, these {\it d}-states take part in two different types of bonding, namely
$\sigma$-type V-V and $\pi$-type V-O bonding,
which leads to two different bandwidths for t$_{2g}$ and e$_g^\sigma$ states.
Since both the metal-metal and  metal-oxygen bonds are mediated through the 
same orbital, a simple analysis of the partial DOS would not allow to distinguish the 
different roles played by the t$_{2g}$ orbitals.
Eyert~{\it et. al.}~\cite{Eyert99} used a local coordinate system
with the Z-axis along the trigonal direction (111) pointing towards the center
of the tetrahedron formed by V ions (see Fig. \ref{tetra})
and plotted the V $d_{3z^2-r^2}$, $d_{xz}+d_{yz}$ and $d_{x^2-y^2}+d_{xy}$
partial DOS.
While $d_{3z^2-r^2}$ orbitals are of pure t$_{2g}$ character,
the other four orbitals comprise a mixture of t$_{2g}$ and e$_g^\sigma$ states.
We have calculated the partial DOS for  a$_{1g}$ and e$_g^\pi$ orbitals, 
using the irreducible representations of the $D_{3d}$ group according to
Terakura~{et al.}\cite{Terakura}
with the following linear combinations of the t$_{2g}$ cubic harmonics:
 the a$_{1g}$ orbital is given by
$(xy+xz+yz)/\sqrt{3}$ and the two e$_g^\pi$ orbitals by $(zx-yz)/\sqrt{2}$
and $(yz+zx-2xy)/\sqrt{6}$.
These three particular linear combinations are valid
if the coordinate axes X, Y and Z are directed along V-O bonds. If
the {\it Z} direction is chosen along one of the trigonal axes
described above,  then
the a$_{1g}$ orbital is the $3z^2-r^2$ orbital in the local coordinate system.

The projected partial LDA DOS of the a$_{1g}$ and the e$_g^\pi$ orbitals
are shown in~Fig.~\ref{dos_V_t2g}.
The bandwidth of the a$_{1g}$ orbital W$_{a_{1g}}$=1.35~eV
is almost a factor of two smaller than the e$_g^\pi$ bandwidth W$_{e_g^\pi}$=2.05~eV.
Nonetheless, we found within LDA for all three t$_{2g}$ derived orbitals
 nearly the same occupancies: n$_{a_{1g}}$=0.60 and n$_{e_{g}^\pi}$=0.45.
Almost all spectral weight
of the a$_{1g}$ orbital is concentrated around the Fermi level in the region
from -0.2~eV to 0.3~eV. In contrast to the a$_{1g}$ orbital the e$_g^\pi$
DOS is flat at the Fermi level. The largest part of the spectral weight
of the  e$_g^\pi$ orbitals is situated in the interval from 0.3 eV to 0.85 eV.
Despite such a different spectral weight the a$_{1g}$ and e$_g^\pi$ bands are
not completely separated in energy.
However, there is a significant difference in the centers of gravity
calculated from the corresponding partial DOS, which can be 
interpreted as a measure for the trigonal splitting of the t$_{2g}$ states.

The trigonal splitting is much smaller than the bandwidth but
has a great importance for the understanding of the physics of the LiV$_2$O$_4$ system
in the presence of  strong Coulomb interaction.
The value and the sign of the trigonal splitting
will determine the orbital in which  the V 3$d$ 
electrons should be localized when a strong Coulomb interaction,
which is larger than the bandwidth, is taken into account.\cite{Anisimov99}
We found that this trigonal splitting value is very sensitive
to the accuracy of the band structure calculations.
In order to increase the accuracy, the overlap between atomic spheres
in our LMTO calculation was set to zero and 
more empty spheres were introduced.
For the following we {\em define} the trigonal splitting of the t$_{2g}$ states
as the difference of the centers of gravity of the a$_{1g}$ and the e$_g^\pi$
projected DOS.
We find that the a$_{1g}$ center of gravity
is 0.26 eV lower than that of the e$_g^\pi$
band~($c_{a_{1g}}$=0.14 eV, $c_{e_g^\pi}$=0.40 eV; 
see Fig.~\ref{dos_V_t2g}).
 One can thus conclude from our
LDA calculations that the a$_{1g}$ orbital is more favorable for the localization
of electrons.

The effective mass $m^*$ is known to be a measure for the Coulomb correlations
and can be obtained  from the electronic specific heat coefficient 
$\gamma$ at low temperatures.
From the LDA-calculated electronic specific heat coefficient $\gamma^{LDA}$,
one can infer the ratio of the effective mass 
to the band mass $m^*$/$m_b$=$\gamma$/$\gamma^{LDA}$.
Here, the former is related to the LDA
DOS at the Fermi level via $\gamma^{LDA}$=$\pi^2k_{B}^2N_AD(E_F)$/3, and
$\gamma$ is the experimental value of the electronic specific heat coefficient
taken from.\cite{Kondo7}
For LiV$_2$O$_4$ we found $m^*$/$m_b$$\approx$~25.8, which is in good agreement with 
previous results.\cite{Mattheiss} Such a huge enhancement of the quasi particle mass
$m^*$ is a strong evidence that Coulomb correlations
are important  in LiV$_2$O$_4$ and have to be taken into account in order to describe
the physics of this system.

\section{A microscopic theory for LiV$_2$O$_4$}
\label{ldadmft_res}

\subsection{LDA+DMFT(QMC) scheme}
\label{ldadmft_scheme}
 
Based on the $d=\infty$ limit\cite{MetzVoll89}, the dynamical mean-field theory 
(DMFT)\cite{vollha93,pruschke,georges96} was developed
as a non-perturbative approach to describe strongly correlated
electron systems. It permits the
calculation of electronic spectra for systems with local electronic Coulomb correlations.
The LDA+DMFT approach is a merger of the
DFT/LDA and the DMFT techniques.\cite{poter97,Lichtenstein98}
It combines the strength of the DFT/LDA, viz describing the weakly correlated
part of the {\em ab initio} Hamiltonian, i.e., electrons in $s$- and $p$-orbitals and
the long-range part of the Coulomb interaction, with the ability 
of the DMFT to treat electronic correlations induced by the local Coulomb
interaction. In this paper we will only briefly discuss the relevant parts
of the LDA+DMFT approach and refer the reader to a recent report
by Held {\it et al.}~\cite{Held01} for more details.

For a given material, one can extract a
LDA Hamiltonian $\hat{H}_{\rm LDA}^{0}$ and
supplement it the local Coulomb interactions:
\begin{eqnarray}
\hat{H} =\hat{H}_{\rm LDA}^{0}&+&U\sum_{m}\sum_{i}
\hat{n}_{im\uparrow}\hat{n}_{im\downarrow }\label{H} \\ &+&
\;\sum_i\sum_{m\neq m'}\sum_{\sigma \sigma'}\;
(U'-\delta_{\sigma \sigma'}J)\;
\hat{n}_{im\sigma}\hat{n}_{im'\sigma'}.
\nonumber
\end{eqnarray}
Here, the index $i$ enumerates the V sites, $m$ denotes
the individual t$_{2g}$  orbitals, and
$\sigma$ the spin.
$H_{{\rm LDA}}^{0}$ is a one-particle Hamiltonian generated from the LDA band
structure with an averaged Coulomb interaction subtracted to avoid a
double counting of the Coulomnb interaction.\cite{poter97}
$U$ is the local intra-orbital Coulomb repulsion
and $J$ the exchange interaction. The local inter-orbital Coulomb repulsion $U^\prime$
is then fixed by rotational invariance: $U^\prime=U-2J$. 
The actual values for
$U$ and $U^\prime$ can be obtained from an averaged Coulomb parameter $\bar U$
and Hund's exchange $J$, which can be calculated with LDA.
The quantity $\bar U$ is related to the Coulomb parameters $U$ and $U'$ via
\begin{equation}
\bar U = \frac{U+(N_{\rm orb}-1)U'+(N_{\rm orb}-1)(U'-J)}{2N_{\rm orb}-1},
\label{ubar}
\end{equation}
where N$_{\rm orb}$ is the number of interacting orbitals (N$_{\rm orb}$=3 in our case).
Since $U$ and $U^\prime$ are
not independent, the two values $\bar{U}$ and $J$ are sufficient to determine
$U$ from this relation.\cite{Held01,Zoelfl00}

The DMFT maps the lattice problem (\ref{H}) onto an effective, self-consistent
impurity problem.  
A reliable method to solve this (multi-band) quantum impurity problem is 
provided by quantum Monte Carlo (QMC) simulations,\cite{QMC}
which are combined with the maximum entropy method~\cite{MEM} for the
calculation of  spectral functions from the imaginary time QMC data.
This technique has been applied to calculate properties of several 
transition metal oxides.\cite{Liebsch00,Nekrasov00,Held01a}

A computationally important simplification is due to the fact that
in cubic spinel the t$_{2g}$ states do not mix
with the e$_g^\sigma$ states. In this particular case the self-energy
matrix $\Sigma_{m\sigma}(z)$ is diagonal with respect to the orbital and spin indices.
Under this condition the Green functions $G_{m\sigma}(z)$ of the lattice
problem can be expressed in the DMFT as Hilbert transform of the non-interacting DOS
$N^0_m(\epsilon)$,
\begin{equation}
G_{m\sigma}(z)=\int d\epsilon
\frac{N^{0}_m(\epsilon )}{z-\mu_{\rm B}B\sigma-\Sigma_{m\sigma}(z)-\epsilon}.
\label{intg}
\end{equation}
This procedure avoids the rather cumbersome and problematic $k$-integration
over the Brillouin zone by the analytical tetrahedron method.\cite{Lambin84}
The LDA DOS $N^{(0)}_m(\epsilon)$ for different orbitals (different $m$) of the 3$d$ t$_{2g}$ states 
in LiV$_2$O$_4$ is displayed in Fig.~\ref{dos_V_t2g}.

A particularly interesting quantity for LiV$_2$O$_4$ is the magnetic susceptibility
$\chi(T)$. In order to calculate $\chi(T)$ we use the definition $\chi(T)=
\lim_{B\to0}M(T)/B$, where $M(T)$ is the magnetization  due to the
applied magnetic field $B$.
Obviously, $M(T)$ can be obtained directly from the QMC data via
\begin{equation}
 \label{magnetization}
 M(T)=\mu_{\rm B}
\sum\limits_{m=1}^{N_{\rm orb}}(\langle n_{m\uparrow}\rangle-\langle n_{m\downarrow}\rangle)
\end{equation}
and
\begin{equation}
\langle n_{m\sigma}\rangle=G_{m\sigma}(\tau=0^+)\;\;.
\end{equation}
Since QMC results are subject to both statistical and systematic errors, a direct
evaluation of $M(T)/B$ for small $B$ will inevitably lead to unpredictable
scattering of the results for $\chi(T)$. To avoid these problems, we perform
calculations for $M(T,B)$ for a series of small fields $B$ and extract $\chi(T)$
from a least-squares fit as the slope of $M(T,B)$ as $B\to0$.

\subsection{Single-particle properties} 
\label{ldadmft_disc}

The DMFT calculations are based on the DFT/LDA-DOS $N^0_m(\epsilon)$
for the a$_{1g}$ and e$_g^\pi$ orbitals
presented in Fig.~\ref{dos_V_t2g}. The total number of electrons
in these three orbitals was fixed to n=1.5, in accordance with the 
+3.5 valency of V ions in LiV$_2$O$_4$.
The $d$-$d$ Coulomb interaction parameter $\bar U$=3.0 eV and exchange
\begin{figure}[htb]
\centering
\epsfig{file=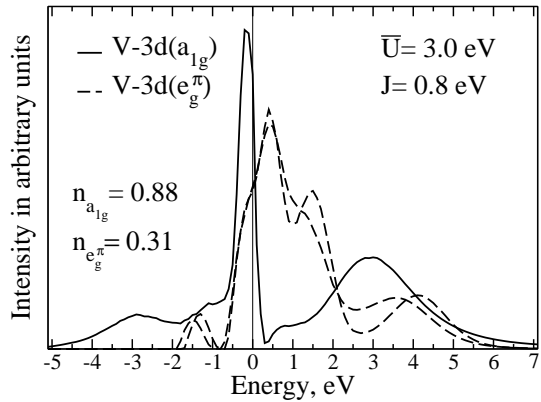,angle=270,width=0.47\textwidth}
\caption{Partial a$_{1g}$ and e$_{g}^\pi$ DOS for LiV$_2$O$_4$
calculated with LDA+DMFT(QMC) for
$\bar U=3.0\,$eV and $J=0.8\,$eV. The non-interacting DOS N$_m^0(\epsilon)$
used in the LDA+DMFT calculations
are presented in Fig.~\ref{dos_V_t2g}.
The LDA+DMFT(QMC) occupancies of the a$_{1g}$ and e$_{g}^\pi$ orbitals
are n$_{a_{1g}}$=0.88 and n$_{e_{g}^\pi}$=0.31.
The Fermi level corresponds to zero energy.}
\label{QMC_0_ovlp}
\end{figure}
Coulomb interaction parameter $J$=0.8 eV were calculated~\cite{Anisimov99}
by the constrained LDA method.\cite{Gunnarsson89}
The temperature used in our QMC simulations was approximately 750~K. While the
scheme in principle poses no restrictions on the temperature values,
the QMC code used presently limits our calculations to these rather high
temperatures because of computing power limitations.\cite{Nekrasov00}

\begin{figure}[htb]
\centering
\epsfig{file=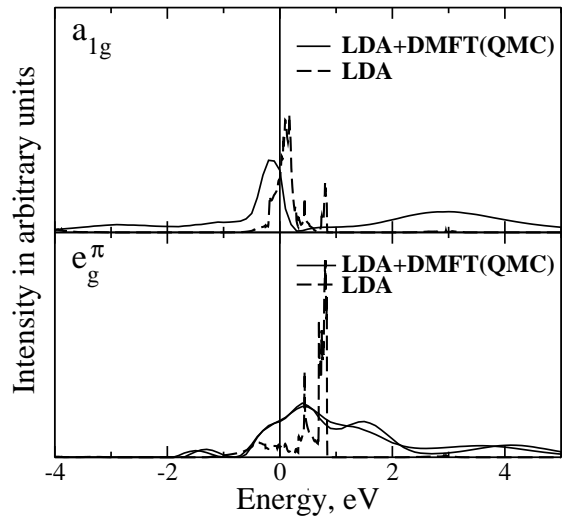,angle=270,width=0.54\textwidth}
\caption{LDA+DMFT(QMC) calculated DOS for LiV$_2$O$_4$
(solid line; $\bar U$=3.0 eV and J=0.8 eV) in 
comparison with non-interacting LDA a$_{1g}$ and e$_g^\pi$ DOS (dashed line).
Please note that for a more convenient
comparison the LDA+DMFT(QMC) DOS were magnified in intensity by a factor of 5.
The Fermi level corresponds to zero energy.} 
\label{dmft_lda}
\end{figure}

The partial DOS of a$_{1g}$ and e$_g^\pi$ orbitals obtained from the analytical
continuation of the QMC results are shown in Fig.~\ref{QMC_0_ovlp}.
In comparison with the
non-interacting case (see Fig.~\ref{dmft_lda}), a considerable transfer of
spectral weight has taken place, especially the  a$_{1g}$ orbital 
appears to be strongly renormalized by Coulomb interactions.
Furthermore, in contrast to the LDA results, where the occupation of 
the a$_{1g}$ and e$_g^\pi$ orbitals is roughly the same, the DMFT yields 
$n_{a_{1g}}\approx0.9$ and $n_{e_{g}^\pi}\approx0.3$. These numbers clearly
demonstrate that one electron is nearly localized in the a$_{1g}$ orbital,
while the e$_g^\pi$ states remain weakly correlated
and metallic with a filling close to 1/8.
Nevertheless, the e$_g^\pi$ orbitals are remarkably broadened
by Coulomb interactions in comparison with the LDA picture (fig.~\ref{dmft_lda}).
Slight differences between the two degenerate e$_g^\pi$ orbitals originate
from the statistical nature of the QMC method and the following
analytical continuation withh MEM.

\begin{table}[htb]
\begin{center}
\begin{tabular}{|lll|l|}
\hline
a$_{1g}$         & e$_g^1$           & e$_g^2$          & \multicolumn{1}{|c|}{Energy}\\
\hline
$|0\rangle$      & $|0\rangle$       & $|0\rangle$      & $0$                                                            \\
$|\sigma\rangle$ & $|0\rangle$       & $|0\rangle$      & $\epsilon_{a_{1g}}$                                            \\
$|0\rangle$      & $|\sigma\rangle$  & $|0\rangle$      & $\epsilon_{a_{1g}}+\Delta\epsilon$                             \\
$|0\rangle$      & $|0\rangle$       & $|\sigma\rangle$ & $\epsilon_{a_{1g}}+\Delta\epsilon$                             \\
$|2\rangle$      & $|0\rangle$       & $|0\rangle$      & $2\epsilon_{a_{1g}}+U$                                         \\
$|\sigma\rangle$ & $|\sigma'\rangle$ & $|0\rangle$      & $2\epsilon_{a_{1g}}+\Delta\epsilon+U'-J\delta_{\sigma\sigma'}$ \\
$|\sigma\rangle$ & $|0\rangle$       & $|\sigma'\rangle$& $2\epsilon_{a_{1g}}+\Delta\epsilon+U'-J\delta_{\sigma\sigma'}$ \\
$|0\rangle$      & $|2\rangle$       & $|0\rangle$      & $2\epsilon_{a_{1g}}+2\Delta\epsilon+U$                         \\
$|0\rangle$      & $|0\rangle$       & $|2\rangle$      & $2\epsilon_{a_{1g}}+2\Delta\epsilon+U$                         \\
$|0\rangle$      & $|\sigma\rangle$  & $|\sigma'\rangle$& $2\epsilon_{a_{1g}}+2\Delta\epsilon+U'-J\delta_{\sigma\sigma'}$\\
\hline
\end{tabular}
\end{center}
\caption[]{Eigenstates and -energies of the atomic Hamiltonian for
total occupations less or equal two.
$\epsilon_{a_{1g}}$ denotes the one particle energy of the a$_{1g}$ orbital
and $\Delta\epsilon\approx0.26$~eV the
trigonal splitting between a$_{1g}$ and e$_g^\pi$ orbitals.\label{tab1}}
\end{table}
To discuss the origin of the structures in the DOS of Fig.~\ref{QMC_0_ovlp},
 it is helpful to look at the spectrum of the atomic Hamiltonian
and the corresponding positions of the one-particle excitations.
The states corresponding to atomic occupancies up to two electrons are listed
in Tab.~\ref{tab1} together with their energies.
Since the Hubbard $U$ is large and we have to accommodate $1.5$ electrons, with one electron
in the a$_{1g}$ orbital, the ground state
has to be a suitable mixture of two states, which both have a singly occupied
a$_{1g}$ orbital, but differ in the
occupancy of the e$_g^\pi$ orbitals. The inspection of all possibilities
in Tab.~\ref{tab1} leaves only
$|\sigma\rangle|0\rangle|0\rangle$, $|\sigma\rangle|\sigma\rangle|0\rangle$
and $|\sigma\rangle|0\rangle|\sigma\rangle$ as
candidates. In order for these three to be (nearly) degenerate,
$\epsilon_{a_{1g}}+\Delta\epsilon+U'-J=0$ or $\epsilon_{a_{1g}}=-U'+J-\Delta\epsilon$.
Inserting the
numbers for $U'$ and $J$ and the LDA-value for $\Delta\epsilon$ leads to
$\epsilon_{a_{1g}}\approx-3.1$~eV. The possible
one-particle excitations with respect to this
ground state configuration and their energies can now be constructed easily,
leading to the results in Tab.~\ref{tab2}.
\begin{table}[htb]
\begin{center}
\begin{tabular}{|l|l|}
\hline
\multicolumn{2}{|c|}{a$_{1g}$ orbital}                                                                       \\
\hline
\multicolumn{1}{|c|}{Excitation process} & \multicolumn{1}{c|}{Excitation energy $\omega_0$}\\ 
\hline
$|\sigma\rangle|0\rangle|0\rangle\to|0\rangle|0\rangle|0\rangle$                   & $\epsilon_{a_{1g}}$     \\
$|\sigma\rangle|\sigma\rangle|0\rangle\to|0\rangle|\sigma\rangle|0\rangle$         & $-\Delta\epsilon$       \\
$|\sigma\rangle|0\rangle|0\rangle\to|2\rangle|0\rangle|0\rangle$                   & $\epsilon_{a_{1g}}+U$   \\
\hline
\multicolumn{2}{|c|}{e$_g^\pi$ orbitals}                                                                       \\
\hline
\multicolumn{1}{|c|}{Excitation process} & \multicolumn{1}{c|}{Excitation energy $\omega_0$}\\ 
\hline
$|\sigma\rangle|0\rangle|0\rangle\to|\sigma\rangle|\sigma\rangle|0\rangle$         & $0$                     \\
$|\sigma\rangle|\sigma\rangle|0\rangle\to|\sigma\rangle|0\rangle|0\rangle$         & $0$                     \\
$|\bar\sigma\rangle|0\rangle|0\rangle\to|\bar\sigma\rangle|\sigma\rangle|0\rangle$ & $J$                     \\
\hline
\end{tabular}
\end{center}
\caption[]{Single-particle excitations for the atomic model filled with $1.5$ electrons.
For spin-degenerate processes
only one representative is listed.\label{tab2}}
\end{table}
It is important to note that the single-particle excitations of the a$_{1g}$
orbital with $\omega_0<0$ have two
distinct contributions, namely one of the usual type ``singly occupied $\to$
unoccupied'', and a second that actually
involves a doubly occupied state build of a mixture of a$_{1g}$ and e$_g^\pi$
states, i.e., a mixed-valent
state. The energy of this latter excitation is given directly by the
trigonal splitting, i.e., this feature can also serve as a means to extract this
number from photoemission experiments.

\begin{figure}[htb]
\begin{center}\mbox{}
\epsfig{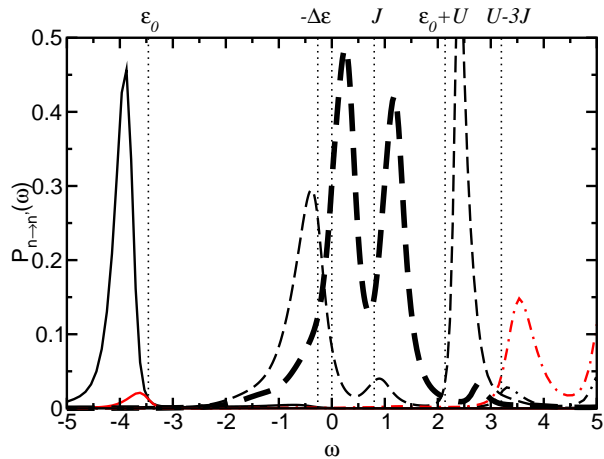} 
\end{center}
\caption[]{One-particle DOS separated into different contributions.
Black (thick)  lines denote a$_{1g}$ (e$_g^\pi$)
single-particle excitations. Full (dashed) lines stand for a singly
occupied (doubly occupied) initial state,
dot-dashed lines for a triply occupied initial state. The dotted
vertical lines show the positions of the excitations
listed in Tab.~\ref{tab2}.}
\label{nca_spec}
\end{figure}

That the excitations listed in Tab.~\ref{tab2} directly map to the peaks in the
DOS Fig.~\ref{QMC_0_ovlp} can be shown by using another
technique to solve the DMFT equations, namely resolvent perturbation theory and
NCA (see e.g.\ Ref.~\onlinecite{Zoelfl00}).
This approach allows for a direct identification of different excitation
channels (i.e., different initial particle numbers)
{\em and} a distinction between a$_{1g}$ and e$_g^\pi$ states. The result is
shown in Fig.~\ref{nca_spec}, where the black lines denote a$_{1g}$
and the gray ones e$_g^\pi$ single-particle excitations.
Full lines stand for a singly occupied
initial state, dashed lines for a doubly occupied initial state and
dot-dashed lines for triply occupied initial
state. Each of the dominant peaks correspond to one of the
transitions listed in Tab.~\ref{tab2}. For clarity, the different
excitation energies are marked by dotted vertical
lines. Apparently, all peaks in Fig.~~\ref{QMC_0_ovlp} have their
counterparts in Fig.~\ref{nca_spec}, explaining them as {\em simply
due to the atomic multiplet structure}.
The shifts are typical renormalizations occurring in quantum impurity models.
Especially the sharp peak at $\omega\approx-0.25$~eV in the a$_{1g}$-DOS
is part of the lower Hubbard band, i.e., the
Fermi energy is located at the upper edge of the lower Hubbard band,
leading to the observed filling close to $1$.
These explanations show that {\em no} Kondo-resonances are seen
in the LDA+DMFT(QMC) spectra
at these elevated temperatures, consistent with
experiment. 
Note also that the structures in the e$_g^\pi$-band around $\omega\approx-1.5$~eV
are related to corresponding features
in the LDA e$_g^\pi$-band located roughly 1.5~eV below the main structure at $0.75$~eV
(see Fig.~\ref{QMC_0_ovlp}).
The double peak structure at $\omega>0$, on the other hand, consists of two
different processes with excitation energies
$\omega_0=0$ and $\omega_0=J$, respectively. Finally, the peak around
$\omega=4$~eV in the e$_g^\pi$ DOS can be identified
as an excitation into a triply occupied state
$|\sigma\rangle|\sigma\rangle\sigma\rangle$ with energy $U-3J$.

The orbital occupation, which confirms the earlier conjecture based on the LDA+U
approach,\cite{Anisimov99} can be readily understood from the previous
analysis of the LDA band structure. There we found that the center of mass of
the a$_{1g}$ orbital is 0.26~eV lower than the corresponding value for
the e$_g^\pi$ orbitals.  In the absence of Coulomb correlations the bandwidth
is significantly larger than this energy difference, i.e., this splitting does
not have any significant effect. 
However, with a Coulomb interaction $\bar U$=3.0 eV, which is significantly larger
than the kinetic energy term (bandwidth $W$$\approx\,$2~eV), this small difference
in connection with the smaller bandwidth of the a$_{1g}$ orbitals will favor a
localization of the electrons in the a$_{1g}$ orbital for energetic reasons.

\subsection{Paramagnetic susceptibility: Competition of two exchange processes}

The LDA+DMFT(QMC) result that roughly one electron is localized in the a$_{1g}$ orbital
has several immediate consequences, which can be tested with experimental
findings. A rather direct consequence is that the electron
localized on the a$_{1g}$ orbital leads to a local moment corresponding to $S=1/2$
per Vanadium atom
and thus a Curie-like susceptibility, consistent with experiment. The remaining
$0.5$ electrons per V ion in the metallic e$_g$ band will lead to a small and temperature
independent Pauli contribution to the susceptibility.

From an experimental point of view, the magnetic properties
of LiV$_2$O$_4$ pose several puzzles. LiV$_2$O$_4$ exhibits a
paramagnetic Curie-Weiss susceptibility\cite{Kondo7,Urano2000,Johnston99,Hayakawa,Kondo99,Onoda97,Lohmann,%
Fujiwara98,Fujiwara99,Mahajan98,Trinkl00,Lee01,Johnston,Takagi87,%
Nakajima91,Johnston95,Ueda97,Kessler71,Chamberland86,Brando02,Miyoshi02} ${\chi(T)= C/(T-\theta) +
\chi_0}$ in the temperature range 50--1000~K.
The best fit
to the experimental data is obtained under the assumption that the
magnetic susceptibility ${\chi(T)}$ is the sum of a Curie-Weiss
and a $T$-independent part $\chi_0$, which contains Pauli
paramagnetic, core diamagnetic and orbital Van Vleck contributions
to the total susceptibility.\cite{Mahajan98} If all V-3$d$
electrons (1.5 per V ion) would equally contribute to the
formation of a local moment, one would expect a mixture of ${S=1}$
and ${S=1/2}$ vanadium ions and consequently an effective paramagnetic
moment 2.34. If only one localized electron in the a$_{1g}$ orbital
contributes to the Curie constant the effective paramagnetic
moment is 1.73.
However, the experimentally observed values of the Curie
constants, as discussed above, in general are close to the latter value of
the effective paramagnetic moment or slightly larger.
Under the assumption that the local spin S is exactly 1/2
one gets a  slightly enchanced g-factor from experiment.
Whereas the negative
Curie-Weiss temperature in the range ${\theta = -20\ldots-60}$~K
indicates antiferromagnetic interactions between local ${S=1/2}$
moments, the large $g$-value points towards ferromagnetic
interactions between local moments and conduction electrons.\cite{Johnston,Johnston95}

A similar ambiguity is revealed by the results of $^7$Li NMR
measurements. The spin--lattice relaxation rate ${1/T_1(T)}$ shows
a broad maximum around 30--50~K and becomes almost temperature
independent above 400~K. The high temperature data (${T \ge
50}$~K) have been interpreted either as an indication of localized
magnetic moments,\cite{Mahajan98} or, on the contrary, as
characteristics of an itinerant electron system close to a
ferromagnetic instability.\cite{Fujiwara98}

\begin{figure}[htb]
\begin{center}\mbox{}
\epsfig{figure=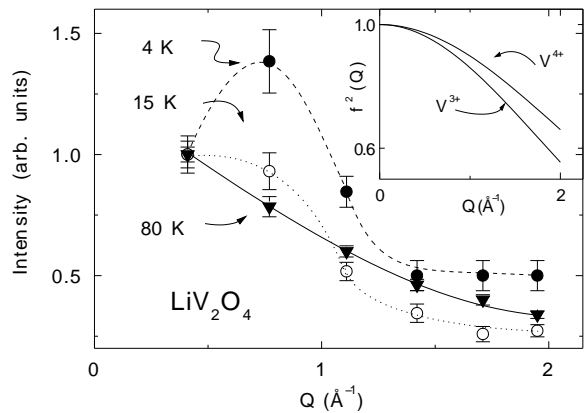,width=0.3\textwidth,angle=-90}
\end{center}
\caption[]{Q-dependence of the energy-integrated quasielastic scattering 
intensity of LiV$_2$O$_4$ at 4, 15, and 80 K. For comparison, the inset shows the Q-dependence
according to the square of the neutron magnetic form factor of the
two Vanadium configurations possible in LiV$_2$O$_4$. At elevated 
temperatures, the much stronger reduction of the 
scattering intensity  upon increasing Q points towards
ferromagnetic spin correlations. The maximum around ${Q=0.7}$
\AA$^{-1}$ at 4 K indicates antiferromagnetic fluctuations.  Dashed or dotted lines are  guides to the eye only (taken from
Ref.~\cite{Krimmel99}).\label{QIntensity}}
\end{figure}

The debate on the relevant magnetic interactions was continued
after the spin fluctuation spectrum of LiV$_2$O$_4$
was  determined by means of quasielastic neutron scattering.\cite{Krimmel99}
These neutron data showed a transition from
ferromagnetic correlations at elevated temperatures (${T \ge
40}$~K)~to antiferromagnetic spin fluctuations with a wave vector
 ${q \approx 0.7}$~\AA$^{-1}$~at low temperatures (${T \le
40}$~K), as shown in Fig.~\ref{QIntensity}. However, a subsequent
inelastic neutron scattering study on LiV$_2$O$_4$ merely revealed
the continuous evolution of antiferromagnetic fluctuations out of
a high-temperature paramagnetic state without indications of
ferromagnetic interactions at elevated temperatures.\cite{Lee01}

In order to clarify these inconsistencies, most recently, a
polarized neutron scattering study on a new set of LiV$_2$O$_4$
samples has been performed.\cite{Kri02}  Full 3-directional
polarization analysis allows for an unambiguous separation of the
nuclear, magnetic and spin incoherent cross-section, respectively.
The measured magnetic cross-section has been fitted and the
corresponding real-space spin correlations extracted by employing
the reverse Monte Carlo method. The data show a temperature
induced cross-over from purely ferromagnetic next nearest neighbor
spin correlations at high temperatures (${T \ge 40}$~K) to a
coexistence of ferromagnetic (nearest neighbor) and
antiferromagnetic (second nearest neighbor) correlations at low
temperatures. The corresponding oscillatory behavior of the
real-space spin correlations at low temperature can be described purely
phenomenologically assuming a RKKY-type interaction
with S(S+1)=0.2 and ${k_F=0.45}$~\AA$^{-1}$~for a simple parabolic
band.\cite{Kri02} Let us estimate the band filling {\em per Vanadium ion}
that this particular value for $k_F$ corresponds to. For a parabolic band
we have
$$
\frac{N_e}{N_{\rm sites}}=n=\frac{1}{N_{\rm sites}}\sum\limits_{k<k_F,\sigma}=2\cdot
\left(\frac{a}{2\pi}\right)^3\cdot\frac{4\pi k_F^3}{3}\;\;,
$$
where $a=8.227$\AA{} is the lattice constant. For $k_F=0.45$\AA$^{-1}$ we obtain
$n\approx1.714$ per cubic unit cell, i.e., a conduction band filling of $0.43$ per
Vanadium ion, since there are four Vanadium ions per unit cell. Note that
this value perfectly coincides with the filling of the e$_g^\pi$ band as
obtained within our LDA/DMFT calculation, thus giving further experimental
confirmation of our proposed picture of a separation of the Vanadium $d$-states
into a strongly correlated a$_{1g}$ orbital with one localized electron
and weakly localized e$_g^\pi$ orbitals which form a metallic band filled with
$\approx0.5$ electrons.

The evolution of antiferromagnetic spin correlations with ${q
\approx 0.7}$~\AA$^{-1}$ at low temperatures is now experimentally
well established. However, the temperature-induced cross-over from
ferromagnetic to RKKY-like spin correlations is still unclear at
present.

We performed LDA+DMFT(QMC) calculations of the high-temperature magnetic susceptibility for
\begin{figure}[htb]
\centering
\epsfig{file=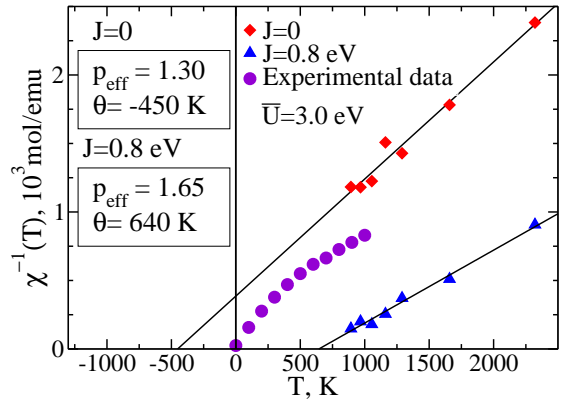,angle=270,width=0.47\textwidth}
\caption{Inverse Curie-Weiss spin susceptibility $\chi^{-1}$(T) for LiV$_2$O$_4$
obtained by LDA+DMFT(QMC) calculation at $\bar U$=3.0~eV,
 $J$=0 (diamonds) and  $J$=0.8~eV(triangles),
in comparison to experimental data\protect\cite{Mahajan98} (circles).
For i) $J$=0 the Curie-Weiss constant is $\theta\approx$-450~K
and the
effective paramagnetic moment is $p_{eff}$=1.30; for
ii) $J$=0.8~eV we obtain $\theta\approx$640~K and $p_{eff}$=1.65.
Solid lines are least squares fits to the LDA+DMFT(QMC) data.
} 
\label{hi}
\end{figure}
a set of different inverse temperatures $\beta$=5, 7, 9, 10, 11, 12, 13~eV$^{-1}$
and magnetic fields $B$=0.005, 0.01, 0.02~eV.
The results are shown as triangles for a Hund's exchange coupling $J$=0.8~eV and 
$J$=0 in Fig.~\ref{hi} together with
a fit to a Curie-Weiss law~\cite{Buczuk02}
$$
\chi(T)=N_A \frac{\mu_{\rm B}^2 p_{eff}^2}{3k_{\rm B}(T-\theta)}\;\;.
$$
The resulting value of the Curie-Weiss temperature is $\theta\approx 640$~K for $J=0.8$~eV, 
the fitted Curie-Weiss constant C gives an effective magnetic moment
$p_{eff}^2=g^2S(S+1)=3.28$, which corresponds
to an effective paramagnetic moment $p_{eff}=1.65$.
The deviation from $p_{eff}=1.73$ for a spin $S=1/2$ can be accounted for if one
recalls that the occupancy of the a$_{1g}$ orbital is not exactly 1 but 0.88.
Such a reduced occupancy should ideally lead to a decrease of $p_{eff}$ from 1.73 to 1.56.
The slightly larger (calculated) value $p_{eff}=1.65$ is due to contributions from the
e$_{g}$ states. For the calculation with $J$=0, where
a$_{1g}$ and e$_{g}$ orbitals do not couple via a local exchange interaction
the value of the effective paramagnetic moment is 1.3, which agrees with
the expectations if one takes into account that for $J$=0 the occupancy of the a$_{1g}$
orbital is 0.75, and hence $p_{eff}=1.73\times0.75=1.3$.

While our value of $p_{eff}$ is in good agreement with known experimental
data, the large ferromagnetic Curie-Weiss temperature of about 640K is
in contrast with experiment. A similar result was obtained by
Anisimov~{\it et~al.},\cite{Anisimov99} where a rather strong
effective ferromagnetic intersite exchange parameter $J_{dex}=530$~K
(which is the sum of direct and double exchanges) was obtained.
A ferromagnetic exchange coupling between local moments in LiV$_2$O$_4$
can be readily understood in the double exchange picture. The presence of
two types of  $d$-electrons, localized ones forming local moments and delocalized
ones producing a partially filled, relatively broad band, is a necessary
ingredient for the double exchange mechanism, resulting in a strong ferromagnetic
coupling between local moments. This requirement is fulfilled in the
case of   LiV$_2$O$_4$ with one electron localized in the a$_{1g}$ orbital
and a 1/8 filled broad  e$_g^\pi$ band. 

A second important condition for the double exchange mechanism is Hund's
intra-atomic exchange, i.e., the ferromagnetic exchange interaction between
electrons within the t$_{2g}$ states.
If one switches off this intra-atomic exchange interaction,
the double exchange mechanism will be switched off, too. This becomes
apparent from the result of our calculations for $\chi^{-1}(T)$ with Hund's
exchange $J$ equal to zero (see Fig.~\ref{hi}, diamonds). Again, we obtain
a Curie-Weiss like behavior with an $p_{eff}$=1.3. This time, however,
the Curie-Weiss temperature $\theta_{dir}\approx-450$~K is {\em negative},
i.e., pointing to an
effective antiferromagnetic exchange. In the absence of intra-atomic exchange and
a$_{1g}$-e$_g^\pi$ hybridization, the only contribution to an exchange coupling
can arise from the
direct hybridization between the a$_{1g}$ electrons. Obviously,
this will result in the observed
antiferromagnetic exchange coupling.

These results clearly show that, neglecting a$_{1g}$-e$_g^\pi$ hybridization,
one has to expect
a subtle competition in  LiV$_2$O$_4$ between {\em antiferromagnetic}\/ direct exchange
resulting from a$_{1g}$-a$_{1g}$ hybridization and {\em ferromagnetic}\/
double exchange from e$_g^\pi$-e$_g^\pi$ hybridization.
Obviously, for the present set of Coulomb parameters the ferromagnetic contribution
wins in our DMFT calculations.
Note, that this result is a direct consequence of the value of $J$ used
in our calculations. There are also available experimental results
from high-energy spectroscopy, which lead one to expect a value of $J=0.65\ldots0.7$~eV
~\cite{exp_J} instead
of $J=0.8$~eV used in our calculation.
We found that such a smaller value of $J$ do not change our results significantly.

There is, however, an important part missing in our present calculation. Since we
use the LDA DOS and not the full Hamiltonian in the DMFT caclulations, the
$a_{1g}$--$e_g^\pi$ {\em hybridization}\/ is neglected completely. As has been
pointed out by  Anisimov~{\it et~al.}\cite{Anisimov99}, this hybridization can give
rise to an antiferromagnetic Kondo coupling between $a_{1g}$ and $e_g^\pi$ orbitals. The value
of this additional exchange coupling was estimated to be $J_K$=-630~K.\cite{Anisimov99}
Thus, in LiV$_2$O$_4$ there are possibly three important exchange interactions
present between the a$_{1g}$ and e$_g^\pi$ electrons of the V3$d$ shell: ferromagnetic
double
exchange ($\approx1090$~K), antiferromagnetic direct exchange ($\approx$-450~K)
and antiferromagnetic Kondo exchange ($\approx$-630~K).
All these interactions effectively cancel, eventually leading to 
a small Curie-Weiss temperature of the order of 0~K in accordance with experiment.

Finally, one puzzle remains, namely that the data from
neutron scattering experiments show ferromagnetic spin
fluctuations at high temperatures.\cite{Krimmel99,Kri02}

The competing exchange terms described above may also explain the
change  from antiferromagnetic to ferromagnetic spin fluctuations at $T\approx 40\,$K
found in neutron scattering experiments.
This is about the same energy scale as the Kondo temperature.
Thus, one might argue that, while below $T_K$ the combined 
antiferromagnetic direct and Kondo exchanges are stronger than
the ferromagnetic double exchange, this changes above $T_K$
where the Kondo effect becomes ineffective, such that
the ferromagnetic exchange prevails.

\section{Conclusion}

We investigated the effect of Coulomb correlations on the electronic
structure, orbital state and magnetic properties of LiV$_2$O$_4$.
The analysis of the non-interacting partial densities of state obtained
by standard LDA calculations shows that, while the trigonal splitting
of  the $t_{2g}$ states into a$_{1g}$ and e$_g^\pi$ orbitals is not strong 
enough to produce separate bands, it leads to a significant difference
in the effective bandwidths and average energy of the trigonal orbitals.
The LDA+DMFT(QMC) calculations gave orbital occupancies and spectra
which indicate a nearly complete localization of one electron out of
1.5 3$d$ electrons per V ion in the a$_{1g}$ orbital, while the e$_g^\pi$ orbitals
form a relatively broad partially filled metallic band.
The calculated temperature dependence of the paramagnetic susceptibility
corresponds to the experimentally observed Curie-Weiss law and
gives an effective paramagnetic moment $p_{eff}$=1.65 in agreement with experimental data.
The experimentally observed small value of the Curie-Weiss temperature, formerly
a puzzle, is supposed to be the result of a competition between three different contributions
to the effective exchange interaction between a$_{1g}$ and e$_g^\pi$  electrons in the V3d shell:
ferromagnetic double
exchange, antiferromagnetic direct exchange
and antiferromagnetic Kondo exchange. The present calculations show a dominance of the first, but
including the hybridization between $a_{1g}$ and $e_g ^\pi$, an almost cancellation of all three
becomes possible leading to the experimentally observed small residual antiferromagnetic
Curie-Weiss temperature.

\begin{acknowledgments}
We thank
A.~Oles,
A.~Loidl,
D.~Singh,
M.~Korotin,
P.~Horsch
and
V.~Eyert
for very useful discussions and
A.~Sandvik
for making available his maximum entropy program.
This work was supported 
in part by the Deutsche Forschungsgemeinschaft through
Sonderforschungsbereich 484 (DV, TP, GK, AK, IN)
and the Russian Foundation for Basic Research through  grants RFFI-01-02-17063 (VA,IN)
and RFFI-02-02-06162 (IN).
\end{acknowledgments}

\end{document}